\documentclass[preprint,preprintnumbers,amsmath,amssymb,superscriptaddress]{revtex4-1}
\usepackage{fullpage,amsthm,amsmath,amsfonts,bm,times,color,bbm}
\usepackage{graphicx,subfigure}
\usepackage[english]{babel}
\usepackage{mathtools}

\def\eqa{\begin{eqnarray}}
\def\eea{\end{eqnarray}}
\newcommand{\eq}{\begin{equation}}
\newcommand{\ee}{\end{equation}}

\renewcommand{\>}{\rangle}

\makeatother

\makeatletter

\begin{document}

\title{Resilience of spatial networks with inter-links behaving as an external field % model and real networks
 }

\author{Jingfang Fan}
%\email{j.fang.fan@gmail.com}
\affiliation{Department of Physics, Bar Ilan University, Ramat Gan 52900, Israel}
\author{Gaogao Dong}
\affiliation{Institute of applied system analysis, Faculty of Science, Jiangsu University, Zhenjiang, 212013 Jiangsu, China}
\affiliation{Energy Development and Environmental Protection Strategy Research Center, Faculty of Science, Jiangsu University, Zhenjiang, 212013 Jiangsu, China}
\affiliation{Center for Polymer Studies and Department of Physics, Boston University, Boston, MA 02215, USA}
\author{Louis M. Shekhtman}
\affiliation{Department of Physics, Bar Ilan University, Ramat Gan 52900, Israel}
\author{Dong Zhou}
\affiliation{Department of Physics, Bar Ilan University, Ramat Gan 52900, Israel}
\author{Jun Meng}
\affiliation{Department of Physics, Bar Ilan University, Ramat Gan 52900, Israel}
\author{Xiaosong Chen}
\affiliation{CAS Key Laboratory of Theoretical Physics, Institute of Theoretical Physics, Chinese Academy of Sciences, Beijing 100190, China}
\affiliation{School of Physical Sciences, University of Chinese Academy of Sciences, Beijing 100049, China}
\author{Shlomo Havlin}
%\email{havlin@ophir.ph.biu.ac.il}
\affiliation{Department of Physics, Bar Ilan University, Ramat Gan 52900, Israel}

\begin{abstract}
Many real systems such as, roads, shipping routes, and infrastructure systems can be modeled based on spatially embedded networks. The inter-links between two distant spatial networks, such as those formed by transcontinental airline flights, play a crucial role in optimizing communication and transportation over such long distances. Still, little is known about  how inter-links affect the resilience of such systems. Here, we develop a framework to study the resilience of interlinked spatially embedded networks based on percolation theory. We find that the inter-links can be regarded as an external field near the percolation phase transition, analogous to a magnetic field in a ferromagnetic-paramagnetic spin system.  
%We also study the universality class 
By defining the analogous critical exponents $\delta$ and $\gamma$, we find that their values for various inter-links structures follow Widom's scaling relations. %DIF >
%DIF -------
Furthermore, we study the optimal robustness of our model and  compare it with the analysis of real-world networks.
The framework presented here not only facilitates the understanding of phase transitions with external fields in complex networks but also provides insight into optimizing real-world infrastructure networks and a magnetic field in a management.
\end{abstract}
\date{\today}

%\flushbottom
\maketitle
\section{INTRODUCTION}
Robustness is of crucial importance in many complex systems and plays an important role in mitigating damage \cite{gao_universal_2016}.
It has been studied widely in both single networks \cite{cohen_resilience_2000,albert2000error,tanizawa2005optimization}, interdependent networks \cite{buldyrev_catastrophic_2010,leicht_percolation_2009,hu_percolation_2011,gao2011robustness,gao_networks_2012,shekhtman2015resilience} and multiplex networks \cite{hackett_bond_2016,sole-ribalta_congestion_2016}. 
Percolation theory has demonstrated its great potential as a versatile tool  
for understanding system resilience based on both dynamical and structural properties \cite{aharony2003introduction,bunde2012fractals}, and has been applied to many real systems \cite{saberi2015recent,li2015percolation,meng_percolation_2017}. Recently, a theoretical framework has been developed to study the resilience of communities formed of either Erd\H os-R\'enyi (ER) and Scale-Free networks that have inter-linkes between them using percolation theory \cite{DongPNAS2018}. It has been found that the inter-links affect the percolation phase transition in a manner similar to an external field in a ferromagnetic-paramagnetic spin system. However, many real systems, such as, transportation networks \cite{weiss_global_2018,strano2017scaling}, infrastructure networks \cite{hines2010topological} and others, are spatially embedded and the influence of this feature has not been considered. Here we study how the inter-links (e.g. air flights) between two spatial networks (e.g., countries) affect the overall resilience. Furthermore, we will search for an optimal structure (or most robust point) of our model and consider it in a real transportation system. We will do so by developing a framework to study the resilience of spatial networks with inter-links and by analysing possible optimal structures for our model/s and in real transport systems.

The structure of our paper is as follows: in the next Chapter, we
describe and introduce the model. In Chapter III, the results are presented and discussed. Finally, in Chap. IV a short summary and outlook are provided.

\section{MODEL}

\begin{figure}
\begin{centering}
\includegraphics[width=1.0\linewidth]{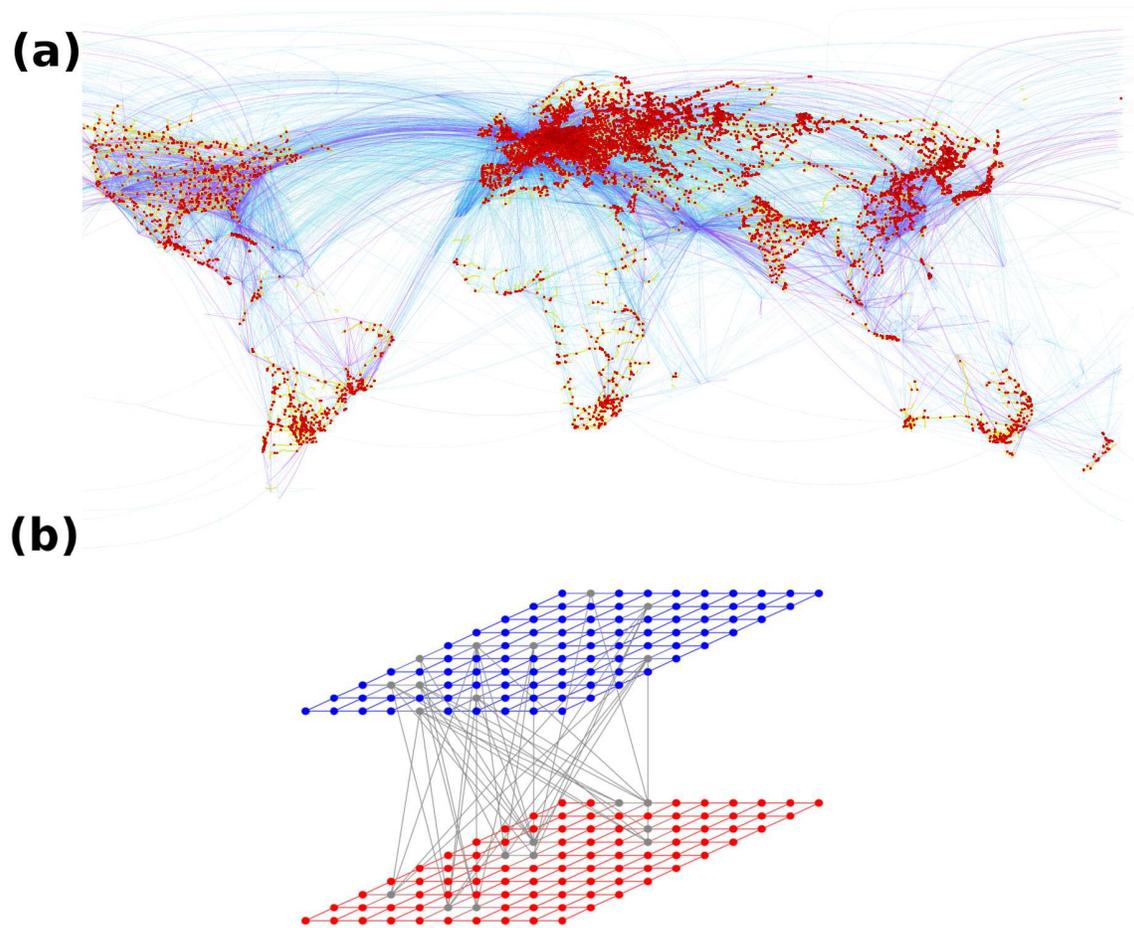}
\caption{\label{Fig_model} (a) The topological structure of the global transport network. The yellow links are railway lines, the red nodes are railway intersections, and the blue lines are global airline routes. (b) Our model. We assume two separate lattice networks, representing two continents (or countries) with railway networks. We add $M_{inter}$ inter-links to a fraction $r$ of nodes, representing cities with airports having flights to the other continent.
Interconnected nodes and their respective interlinks are highlighted in gray. Here, we chose $r$ = 0.1 and $M_{inter}$ = 50.}
\end{centering}
\end{figure}

Our model is motived by many real-world networks where nodes and links are spatially embedded within the same region (module), but only some nodes have connections to other regions (modules). We denote the links in the same module as \textit{intralinks} and the links between different modules as 
\textit{interlinks}. Fig.~\ref{Fig_model}(a) demonstrates the topological structure of the global transportation network including railway roads and airline routes \cite{DATA}. We demonstrate in the figure  that the airports are connected via interlinks and can be regarded as \textit{interconnected nodes}. We show here that the interlinks behave, regarding breakdown of the network, in a manner analogous to an external field from physics near magnetic-paramagnetic phase transition \cite{Stanley_1971,reynolds1977ghost}. To study this effect, for simplicity and without loss of generalization, we carried out extensive simulations on a network of two modules each with the same number of nodes, \textit{$N_{1} = L\times L$}, where $L$ is the linear size of the lattice, representing the spatial networks. Within each module the nodes are only connected with their neighbors in space as defined by a 2-dimensional square lattice. Between different modules,  we randomly select a fraction $r$ of nodes to be interconnected nodes, e.g, airports, and randomly assign $M_{inter}$ interlinks among nodes in the two modules.  A network generated from our model is shown in Fig.~\ref{Fig_model}(b). Our model is realistic and can represent coupled transport systems, i.e, the nodes in the same lattice module are localized railroad or road networks within the same region while the interlinks represent interregional airline routes.

To quantify the resilience of our model, we carried out extensive numerical simulations of the size of the giant connected component $S(p,r)$ after  a fraction of $1 - p$ nodes are randomly removed. Note that our model is distinct from the case of interdependent networks \cite{buldyrev_catastrophic_2010}, where the failure of nodes in one network leads to the failure of dependent nodes in other networks. 
Our model is also different from the interconnected modules model \cite{shai_critical_2015}, where interconnected nodes are attacked.
In our model, the interconnections between different communities are additional connectivity links \cite{li2014epidemics} and randomly chosen nodes are attacked \cite{DongPNAS2018}.
For a given set of parameters $[p,r;L]$, we carried out 10,000 Monte Carlo realizations and took the average of these results to obtain $S(p,r)$.

\section{RESULTS}

\begin{figure}
\begin{centering}
\includegraphics[width=1.0\linewidth]{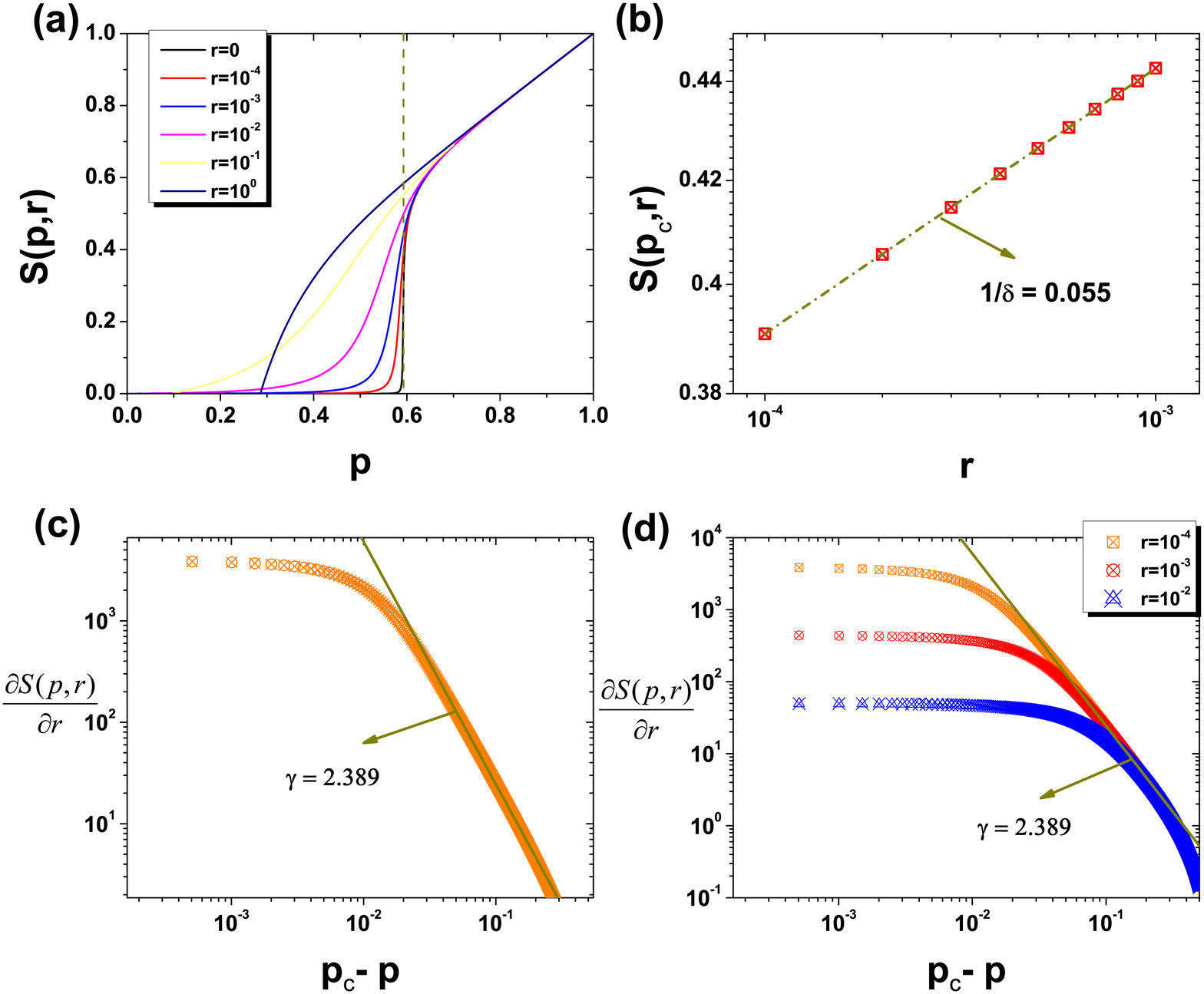}
\caption{\label{Fig:1A} (a) The giant component (order parameter), $S(p,r)$, as a function of the fraction of non-removed nodes $p$ for several values of $r$; (b) $S(p_c,r)$ as a function of $r$ with the exponent $\delta$; (c) $\frac{\partial S(p,r)}{\partial r}$ as a function of $p_{c}-p$ with $r = 10^{-4}$ and the exponent $\gamma$; (d) Same as (c) but for several $r$. Here, $L = 4096$, $M_{inter} = 2\times L\times L$, $p_c = 0.592746$. The dashed line is the best fit-line for the data, which is found to have a slope $1/\delta = 0.055$ and R-Square $>0.999$.}
\end{centering}
\end{figure}

Similar to our earlier studies \cite{DongPNAS2018,shekhtman_critical_2018}, we find that the parameter $r$, governing the fraction of interconnected nodes, has effects analogous to a magnetic field in a spin system, near criticality. This analogy can be seen through the facts that: (i) the non-zero fraction of interconnected nodes destroys the original phase transition point of the single module; (ii) critical exponents (defined below) of values derived from percolation theory can be used to characterize the effect of external field on $S(p,r)$.
Fig.~\ref{Fig:1A}(a) shows our simulation results for the size of the giant component $S(p,r)$ with $L = 4096$,
$M_{inter} = 2\times L\times L$ for various $r$. We note that in the limit of $r=0$ our model recovers the critical threshold of  single square lattices, $p_{c} \approx 0.592746$ \cite{newman_efficient_2000}. We find that $S(p_c,r) >S(p_c,0)=0$ for $r>0$, showing that the interconnected nodes remove the phase transition of the single lattice.

Next, we investigate the scaling relations and critical exponents, with $S(p,r)$, $p$ and $r$ serving as our analogy for magnetization (order parameter), temperature, and the external field respectively \cite{Stanley_1971}.
To quantify how the external field, $r$, affects the phase transition, we define the critical exponents $\delta$, which relates the order parameter at the critical point to the magnitude of the field,
\begin{equation}\label{eq5}
S(p_c,r) \sim r^{1/\delta},
\end{equation}
and $\gamma$, which describes the susceptibility near criticality,
\begin{equation}\label{eq6}
\left (\frac {\partial S(p,r)} {\partial r}  \right)_{r\rightarrow 0} \sim  \left| p - p_c \right|^{-\gamma},
\end{equation}
where $p_c$ is the site percolation threshold for a single 2-dimensional square lattice network.  

The simulation results for $\delta$ in our model are shown in Fig.~\ref{Fig:1A}(b). We obtain $1/\delta = 0.055$ from simulations, which agrees very well with the known exponent value for standard percolation on square lattices $1/\delta = 5/91$ \cite{aharony2003introduction,bunde2012fractals}. %{\color{red} LMS: Didn't we derive this based on $\gamma$ and $\beta$? If it is not explicitly mentioned in the references then we should mention that it is derived using the scaling relations}. 
The dashed line is the best fit-line for the data with R-Square $>0.999$. 

We next investigate the critical exponent, $\gamma$, which we claim to be analogous to magnetic susceptibility exponent with the scaling relation given in Eq.~\eqref{eq6}. Fig.~\ref{Fig:1A}(c) presents our results for $\gamma$. We obtain $\gamma = 2.389$ for $p < p_c$ and $r = 10^{-4}$, which agrees again very well with the known value $\gamma = 43/18$ in percolation \cite{aharony2003introduction,bunde2012fractals}. In Fig.~\ref{Fig:1A}(d) we also plot our results for different $r$ values: $r = 10^{-4}, 10^{-3}, 10^{-2}$ to highlight the changes in the range of the scaling region. We find that as $r$ decreases, the scaling region becomes larger, this is expected since for smaller $r$ the system approaches closer to criticality ($r$=0).
%
%This can be explained as follows: since the inter-links are randomly added between the interconnected nodes, we can assume that $p^{inter}_{c} \sim \frac{1}{\<k_{inter}\>} = \frac{rN}{2M_{inter}}$ \cite{cohen_resilience_2000}.  Thus, smaller $r$ suggests a stronger external field, resulting a larger scaling region. 
Similar effects in terms of the scaling range are also observed for changing $M_{inter}$ with respect to the critical exponent $1/\delta$ and Eq.~\eqref{eq5}, as seen in Fig.~\ref{Fig:S1A} \cite{SI}.

We note that for a single 2d square lattice, the scaling exponent $\beta$, defined by the relation $S \sim (p - p_c)^{\beta}$, has a value of $\beta = 5/36$ \cite{aharony2003introduction,bunde2012fractals}. 
The critical exponent $\beta$ together with $\delta$ and $\gamma$ characterize the percolation universality class for our model. Since the  various thermodynamic quantities are related, these critical exponents are not independent, but rather can be uniquely defined in terms of only two of them \cite{domb2000phase}. We find that the scaling hypothesis is also valid for our model and note that our values for these exponents are consistent with the Widom's identity $\delta -1 = \gamma/\beta$ \cite{bunde2012fractals}.

\begin{figure}
\begin{centering}
\includegraphics[width=1.0\linewidth]{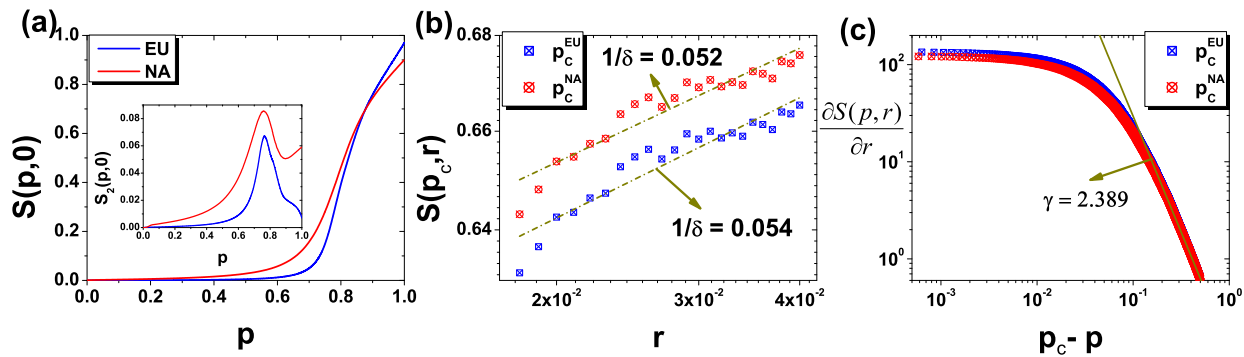}
\caption{\label{Fig:2A} (a) $S(p,0)$, versus the fraction of non-removed nodes, $p$, for real-data of the European (EU) and North America (NA) railway networks; (b) $S(p_c,r)$ as a function of $r$; (c) $\frac{\partial S(p,r)}{\partial r}$ as a function of $p_{c}-p$ for $r =  10^{-2}$. Inset in (a) shows the second largest component $S_{2}(p,0)$ as a function of $p$. We obtain our values of $p_c$ based on the peak of $S_{2}(p,0)$, which gives $p_{c}^{EU} = 0.7641$ and $p_{c}^{NA} = 0.7578$. The dashed lines in (b) are the best-fit  lines for the data with slopes $1/\delta =0.054$, $1/\delta =0.052$ and R-Square $>0.89$. The network sizes are $N_{EU}=8354$, $M_{EU}=11128$; $N_{NA}=933$, $M_{NA}=1273$, $M_{flight} = 1864$.}
\end{centering}
\end{figure}

In the following, we test our framework on a real world example involving global transportation networks. We consider two railway networks, one in Europe (EU) and the other in North America (NA). The two railway networks have $N_{EU}=8354$ and $N_{NA}=933$ nodes (stations), as well as $M_{EU}=11128$ and $M_{NA}=1273$ intralinks respectively. As an example of adding long distance flights, we added $M_{flight}$ interconnected links randomly among $r$ fraction of the nodes (airport hubs). We used $M_{flight} = 1864$, which is the actual number of direct flights between the two continents. Fig. \ref{Fig:2A} shows our results for the system of the two real networks. We find that, the values of the critical exponents $\delta$ and $\gamma$ for the real networks [Fig. \ref{Fig:2A}(b) and (c)] are consistent with the results obtained from our model. One should note that the percolation threshold $p_c$ is different in each module when they are separated, since the number of nodes and links is not the same in both modules. To obtain the percolation threshold, $p_c$ for each real railway network, we analyzed the second largest component, $S_{2} (p,0)$. The size of the second largest cluster is known to be at a maximum at $p_c$ \cite{margolina_size_1982}.
We obtained  $p_{c}^{EU} = 0.764$ and $p_{c}^{NA} = 0.758$  by utilizing the peak of $S_{2} (p,0)$ for the EU and NA networks respectively [see inset of Fig. \ref{Fig:2A}(a)].

%One of the most important parameters for phase transitions is the critical threshold $p_c$, which defines the point of the transition. Despite the importance of $p_c$, there is still no exact solution for site percolation in the 2d square lattice. 

\begin{figure}
\begin{centering}
\includegraphics[width=1.0\linewidth]{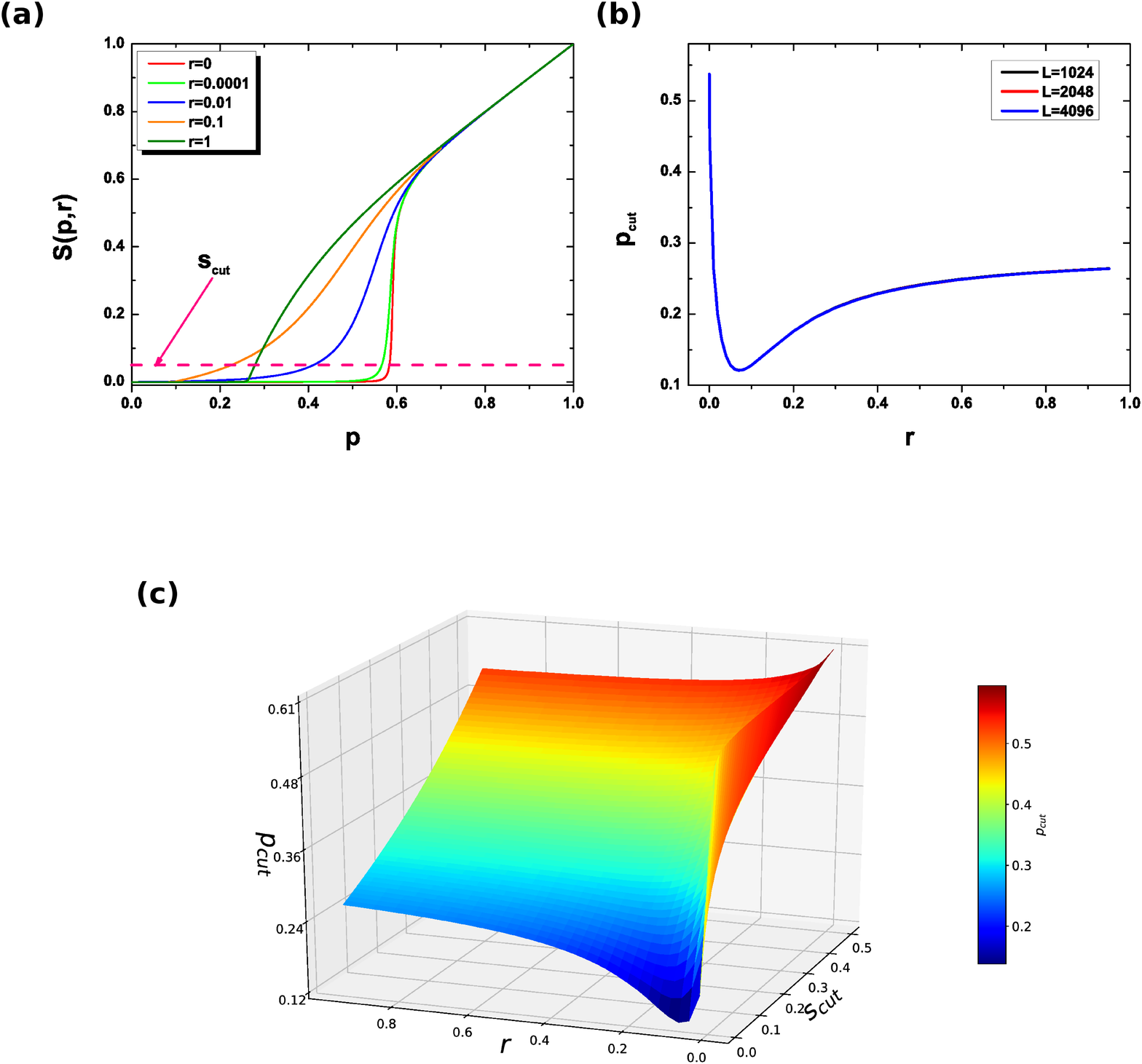}
\caption{\label{Fig:3A} The effective percolation threshold, $p_{cut}$, for our model. (a) Definition of $p_{cut}$ as the intersection between $S(p,r)$ and $S_{cut}$. (b) $p_{cut}$ as a function of $r$ with $S_{cut} = 0.01$. (c) $p_{cut}$ as a function of $r$ and $S_{cut}$. }
\end{centering}
\end{figure}

To analyze the robustness of our model, we define an effective percolation threshold, $p_{cut}$, by using a small cut-off value of the giant component $S_{cut}$, as shown in Fig.~\ref{Fig:3A}(a). The threshold $p_{cut}$ is defined as the point where $S(p,r)$ reaches $S_{cut}$. We assume that when $S(p,r)$ is very small as $S_{cut}$ or below it is not functional. Interestingly, we find an optimal  $r$ in our model. It means that for a certain $r=r_{opt}$ the system is most robust i.e., $p_{cut}$ is minimal. Indeed, Fig.~\ref{Fig:3A}(b) shows a specific example with $S_{cut} = 0.01$, where we find the optimal point to be $r_{opt} \approx 0.05$. In our framework, this suggests that if $5\%$ of the cities have interconnected flights the network is most robust to random failures.  The origin of this optimization phenomenon is due to the percolation competition  between the individual lattice module and the interconnected `network' composed of $r$ interconnected nodes/inter-links. When $r$ is small enough, the behavior of the giant component 
$S(p,r)$ is dominated by the single lattice module [see Fig.~\ref{Fig:3A}(a)],
and the threshold $p_{cut}$ is large and close to $p_c$ [see Fig.~\ref{Fig:3A}(b), with small $r$];
when $r$ is increasing, the effect of the giant component of a single lattice module becomes weaker, but the effect from the interconnected nodes/inter-links becomes stronger resulting the decreasing of $p_{cut}$; however, when $r$ is large, the behavior of the giant component is dominated by the interconnected nodes/inter-links, $p_{cut}$ is proportional to $r$ [see Fig.~\ref{Fig:3A}(b), with large $r$]. In particular, our model will become like a random network, when $r=1$. 
We also find that, in Fig.~\ref{Fig:3A}(b), there are no significant finite-size effects for our system since the three curves with $L = 1024, 2048, 4096$ are nearly overlapping. The results on how $p_{cut}$ changes with $S_{cut}$ and $r$ are shown in Fig.~\ref{Fig:3A}(c).

\begin{figure}
\begin{centering}
\includegraphics[width=1.0\linewidth]{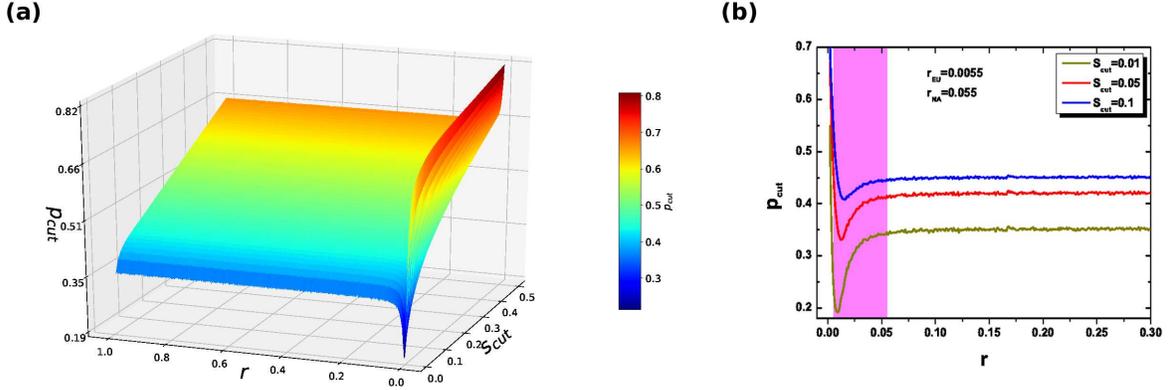}
\caption{\label{Fig:4A} The effective percolation threshold for a real-world network. (a) $p_{cut}$ as a function of $r$ and $S_{cut}$. (b) $p_{cut}$ as a function of $r$ with $S_{cut} = 0.01, 0.05, 0.1$. The region between $r_{EU} = 0.0055$ and $r_{NA} = 0.05$ is highlighted. }
\end{centering}
\end{figure}

Fig.~\ref{Fig:4A}(a) presents how $p_{cut}$ changes with $S_{cut}$ and $r$ for a real network. These results are qualitatively similar to our model results [Fig.~\ref{Fig:3A}(c)]. We also observe that there exists an optimal value of $r$ in the real transportation network.  Fig.~\ref{Fig:4A}(b) shows three specific cases with $S_{cut} = 0.01, 0.05, 0.1$. We find
that the optimal point is around $r_{opt} \approx 0.01$. Suggesting that if $1\%$ of cities have intercontinental flights the system is optimally robust against random failures. For comparison, we also show in the figure the fraction of interconnected nodes in the real data: $r_{EU} = 0.0055$ and $r_{NA} = 0.05$. The lower and upper boundaries of the shadow  in Fig.~\ref{Fig:4A}(b) are based on these two values.

Note that the number of interconnected links, $M_{inter}$, is kept constant when we change $r$ in our model, i.e, $\<k_{inter}\>$ is proportional to $1/r$.
We also performed the same analysis to identify how the external field  affects the resilience, i.e., the critical exponents $\delta, \gamma$ and effective percolation threshold of the spatial and ER networks when $\<k_{inter}\>$ is fixed and $M_{inter}$ changes, according to $\<k_{inter}\> = \<M_{inter}\>/(rN)$. The results are presented and discussed in Supplemental Materials \cite{SI}.

%{\color{red} LMS: The whole above paragraph is unclear to me, and the new models seem to lack motivation. Looking at Figure S2, I have difficulty thinking of the motivation myself. For the Wei Li PRL 2012 paper it made sense to consider distance of interlinks since the infrastructure networks were overlapping, here we are saying they are distance so what is the motivation?}

%\subsection{UNIVERSALITY}
%\subsection{OPTIMIZATION}
\section{SUMMARY}
We have developed a framework to study the resilience of coupled spatial networks where we show that the inter-links act analogously to an external field in a magnetic-paramagnetic system. Using percolation theory we studied the dynamical evolution of the giant component, and found the scaling relations governing the external field. We defined the critical exponents $\delta$ and $\gamma$ using $S$, $p$ and $r$, which serve as analogues of the total magnetization, temperature and external field, respectively.  The values of the critical exponents are universal and relate well with the known values previously obtained for standard percolation on a 2d lattice. Furthermore, we find that our scaling relations obey the Widom's identity. 

We next defined the effective percolation threshold to quantify the robustness of our model. We found that there exists an optimal amount of interconnected nodes, which is also predicted and observed in real-world networks. 
Our approach provides a new perspective on resilience of networks with community structure and gives insight on its interlinks  response as an external field. Lastly, our model provides a method for optimizing real world interconnected infrastructure networks which could be implemented by practitioners in the field.

\section*{Acknowledgements}
We acknowledge the Israel-Italian collaborative
project NECST, the Israel Science Foundation, the Major Program of National Natural Science Foundation of China (Grants 71690242, 91546118), ONR, Japan
Science Foundation, BSF-NSF, and DTRA (Grant no.
HDTRA-1-10-1-0014) for financial support. This
work was partially supported by National Natural Science Foundation
of China (Grants 61403171, 71403105, 2015M581738 and 1501100B) and Key Research Program of Frontier Sciences, CAS, Grant No. QYZDJ-SSW-SYS019. J.F thanks the fellowship program funded by the Planning and Budgeting Committee of the Council for Higher Education of Israel.

\bibliography{MyLibrary}

\appendix

%%%%%%%%%% Merge with supplemental materials %%%%%%%%%%
\pagebreak
\widetext
\begin{center}
\textbf{\large Supplemental Materials: Resilience of spatial networks with inter-links behaving as an external field}

Jingfang Fan, Gaogao Dong, Louis M. Shekhtman, Dong Zhou, Jun Meng, Xiaosong Chen and Shlomo Havlin
\end{center}

\setcounter{equation}{0}
\setcounter{figure}{0}
\setcounter{table}{0}
%\makeatletter
\renewcommand{\theequation}{S\arabic{equation}}
\renewcommand{\thefigure}{S\arabic{figure}}

%\begin{equation}
%\left\{
%\begin{array}{ll}
%\boldsymbol{x}_{<}  =  \boldsymbol{z}_{<},  \\ 
%\boldsymbol{x}_{>}   =  \boldsymbol{z}_{>} \odot e^{s(\boldsymbol{z}_{<})} + t(\boldsymbol{z}_{<}), \label{eq:rnvp}
%\end{array}
%\right.
%\end{equation}
%The transformation \Eq{eq:rnvp} is easy to invert by reversing the basic arithmetical

\section{Further results}
We present here some further results not given in the main text.

\begin{figure}
\begin{centering}
\includegraphics[width=1.0\linewidth]{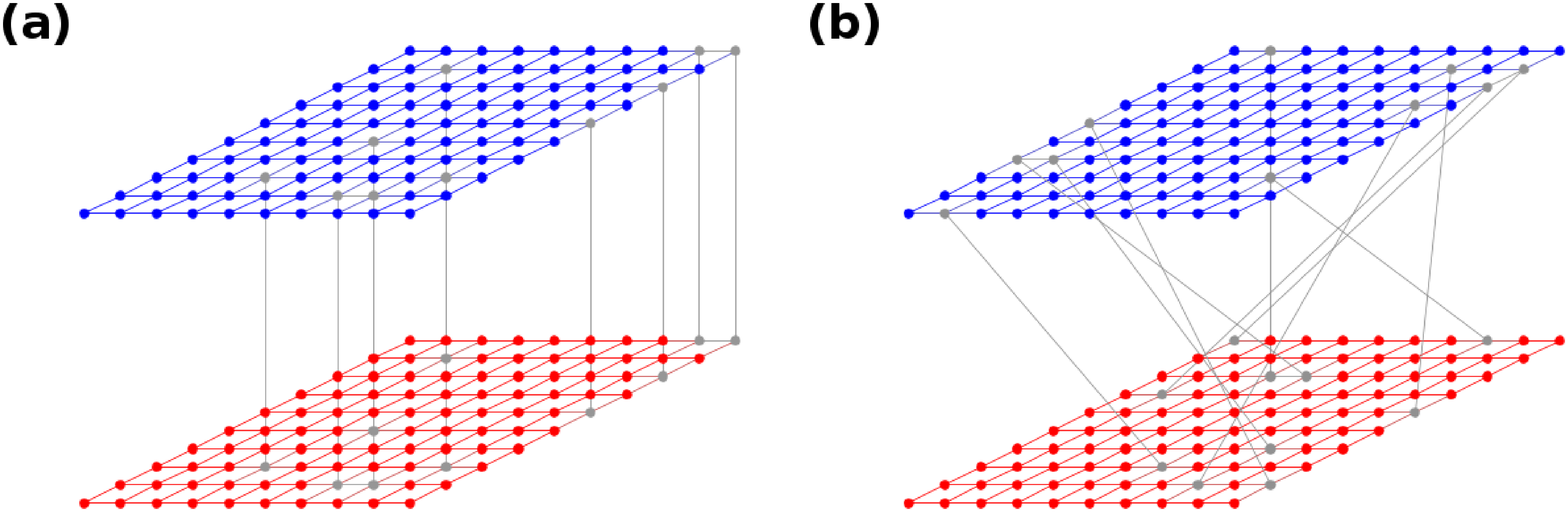}
\caption{\label{Fig:S1} Our model with $\<k_{inter}\> = 1$ and $r = 0.1$ for (a) $d=0$ and (b) $d<\infty$.}
\end{centering}
\end{figure}

\begin{figure}
\begin{centering}
\includegraphics[width=1.0\linewidth]{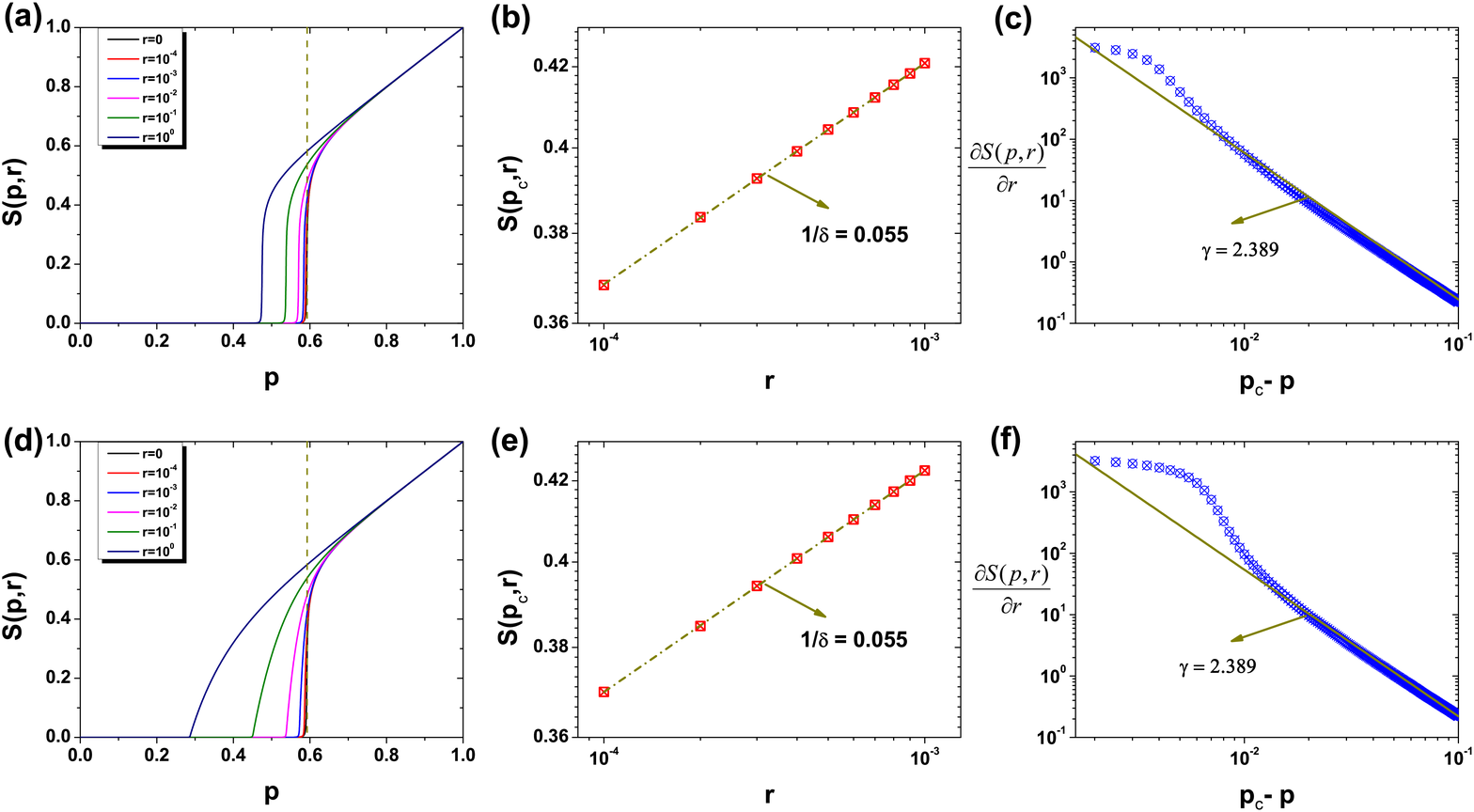}
\caption{\label{Fig:S2} Similar to Fig.~\ref{Fig:1A} from the main text but for the additional models with $\<k_{inter}\> = 1$ and 
(a)(b)(c)  the distance $d=0$, (d)(e)(f) $d<\infty$.}
\end{centering}
\end{figure}

\begin{figure}
\begin{centering}
\includegraphics[width=1.0\linewidth]{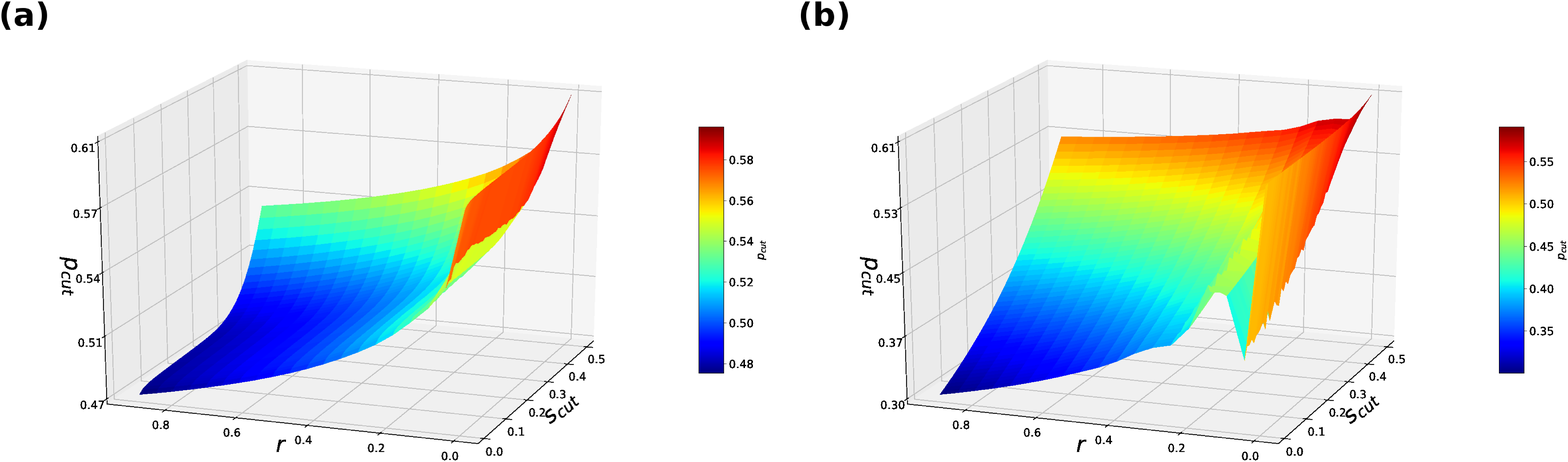}
\caption{\label{Fig:S3} Effective percolation threshold for the additional models with $\<k_{inter}\> = 1$ and 
(a) $d=0$, (b) $d<\infty$.}
\end{centering}
\end{figure}

We consider two additional models with $\<k_{inter}\> =1$.  For these models we randomly add $M_{inter} = rN_{1}$ inter-links between two lattice networks  
within a distance (i) $d=0$ [as shown in Fig.~\ref{Fig:S1}(a)] and (ii) $d<\infty$  [as shown in Fig.~\ref{Fig:S1}(b)],
the definition  of $d$ can refer in \cite{li_cascading_2012}. We repeat our analysis in these two new models and present the results in Fig.~\ref{Fig:S2}.
We find that the critical exponents $\delta$ and $\gamma$ are constant for these cases and do not change with  $\<k_{inter}\>$ and $d$.
However, the optimization phenomenon is absent in both new models in terms of the effective percolation threshold [in Fig.~\ref{Fig:S3}].

\begin{figure}
\begin{centering}
\includegraphics[width=1.0\linewidth]{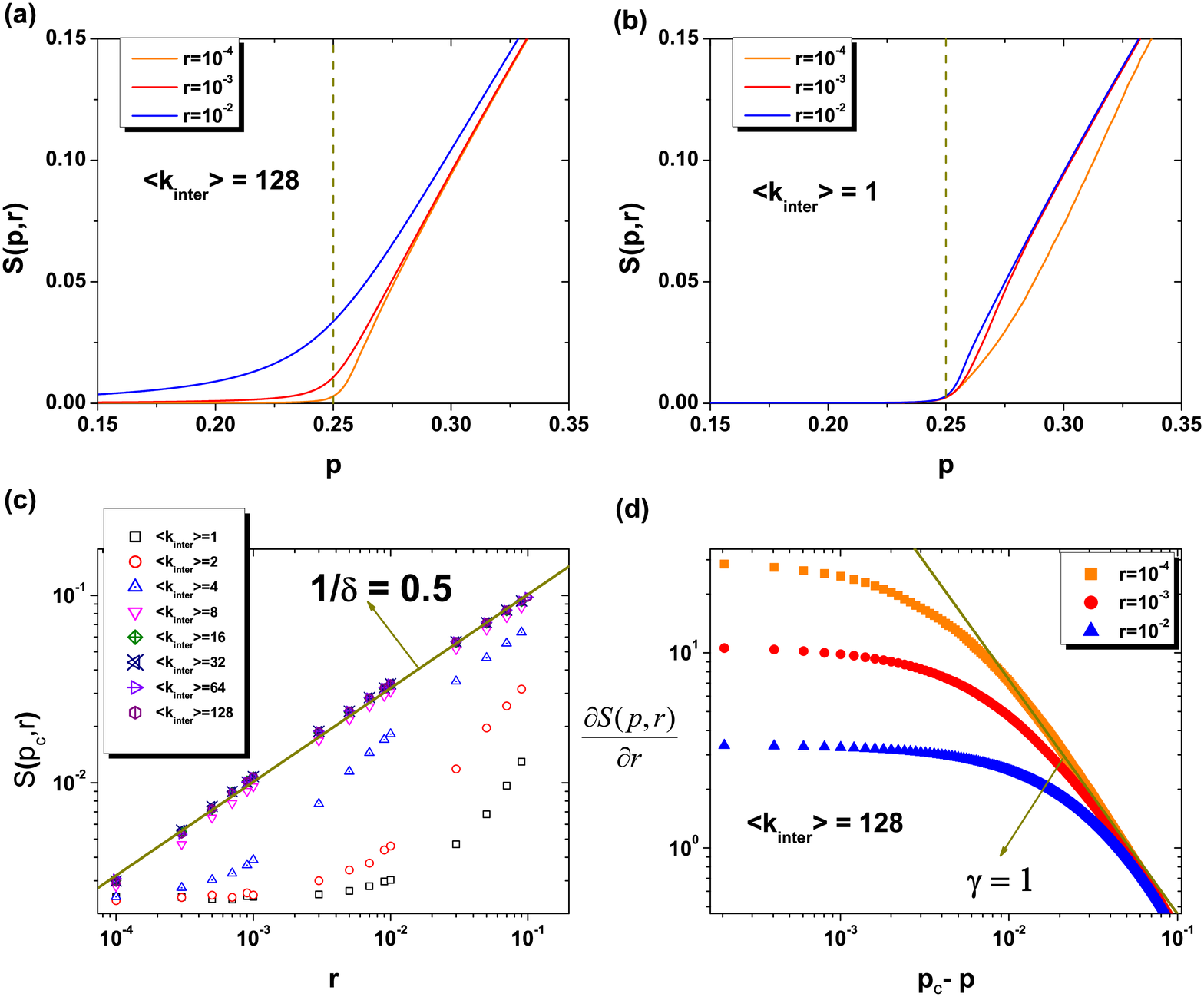}
\caption{\label{Fig:S4} The size of the giant component $S(p,r)$ of two interconnected ER networks as a function of $p$ for varying $r$ with: (a)  $\<k_{inter}\> =128$ and (b) $\<k_{inter}\> =1$.  (c) $S(p_c,r)$ as a function of $r$ for varying $\<k_{inter}\>$. (d) $\frac{\partial S(p,r)}{\partial r}$ as a function of $p_{c}-p$ with $\<k_{inter}\> =128$. The average degree for each individual ER
network is $\<k\>=4$ for all plots. The vertical dashed lines in (a) and (b) show the critical threshold for a single ER network, $p_c = 1/\<k\>=0.25$ . The solid green lines in (c) and (d) show the slope $1/\delta = 0.5$ and $\gamma = 1$, respectively.}
\end{centering}
\end{figure}

In addition, we also study the external field effect on two ER networks with a fixed $\<k_{inter}\>$. Distinct from our models, we find that the value of $\<k_{inter}\>$ significantly influences the critical exponents: only for large $\<k_{inter}\>$ are Eqs.~\eqref{eq5} and \eqref{eq6} satisfied with the mean-field values $\delta = 2$, $\gamma =1$ \cite{DongPNAS2018} [see Fig.~\ref{Fig:S4}]. The origin of such difference is that,
%the competition percolation between resilience of the individual module and
%the interconnected `network' composed of $r$ interconnected nodes and inter-links.
%{\color{red} LMS: The previous sentence is unclear to me. Please check if my revisions are correct, but even still I think more detail is needed.} 
for smaller  $\<k_{inter}\>$, the external field is not strong enough to distinguish the percolation threshold for different $r$ [see Fig.~\ref{Fig:S4}(b)]. In contrast it is easy to distinguish the changes from varying $r$ with large $\<k_{inter}\>$ [see the Fig.~\ref{Fig:S4}(a)]. 

\clearpage

\begin{figure}
\begin{centering}
\includegraphics[width=1.0\linewidth]{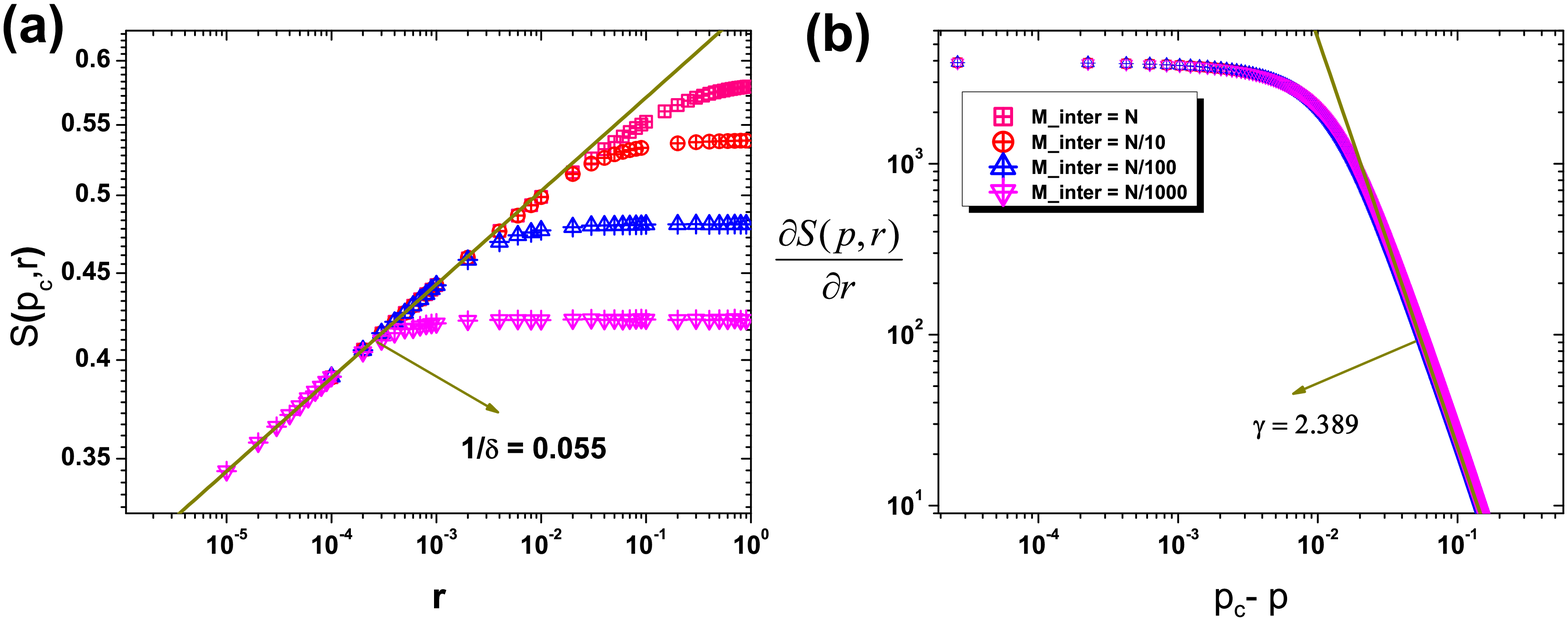}
\caption{\label{Fig:S1A} (a) $S(p_c,r)$ as a function of $r$ for different $M_{inter}$. (b) $\frac{\partial S(r,p)}{\partial r}$ as a function of $p_{c}-p$ for $r = 10^{-4}$ and different $M_{inter}$. The solid green lines in (a) and (b) show the slope $1/\delta = 0.055$ and $\gamma = 2.389$ respectively. }
\end{centering}
\end{figure}

\end{document}